# Spin-Directed Momentum Transfers in SIDIS Baryon Production

Dennis Sivers
Portland Physics Institute
University of Michigan

Abstract

The measurement of transverse single-spin asymmetries for baryon production in the target fragmentation region of semi-inclusive deep-inelastic scattering (SIDIS), can produce important insight into those nonperturbative aspects of QCD directly associated with confinement and with the dynamical breaking of chiral symmetry. We discuss here, in terms of spin-directed momentum transfers, the powerful quantum field-theoretical constraints on the spin-orbit dynamics underlying these transverse spin observables. The $A_\tau$-odd spin-directed momentum shifts, originating either in the target nucleon $(\delta k_{TN})$ or in the QCD jets $(\delta p_{TN})$ produced in the deep inelastic scattering process, represent significant quantum entanglement effects connecting information from current fragmentation with observables in target fragmentation.

Instrumentation of the "target fragmentation" region at an electron-ion collider (EIC) can provide a broad range of new tools for the study of nonperturbative dynamics in QCD. [1,2] The specific observables discussed here are the transverse single-spin asymmetry measurements

$$A_N d\sigma(ep\uparrow \Rightarrow e'BX) \quad B = p, n, \Lambda, \Sigma, \Delta...$$
$$P d\sigma(ep \Rightarrow e'B\uparrow X) \quad B\uparrow = \Lambda\uparrow, \Sigma\uparrow, \Xi\uparrow... \qquad (1.1)$$

The systematic, organized program of such measurements should also be done at the Jefferson Lab CEBAF 12 GeV machine [3] due to the importance of single-spin observables in understanding spin-orbit dynamics in a full range of kinematics. As described below, these baryon asymmetry measurements both supplement and enhance the understanding of measurements of target spin asymmetries for meson production

$$A_N d\sigma(ep\uparrow \Rightarrow e'MX) \quad M = \pi^\pm, \pi^0, K^\pm, K^0, \eta^0... \qquad (1.2)$$

found in the "current fragmentation" region. It is convenient to parameterize the baryon observables of (1.1) in terms of the transverse-momentum dependent (TMD) "fractured functions" introduced in references [1] and [2]. For the target-spin asymmetries we can define the Fractured Orbital Functions (FOF) $\Delta^N M^q_{B/(q,q):p\uparrow}(x,k_{TN};z,\vec{p}^B_T\cdot\vec{k}_T;Q^2)$, and the Fractured Collins, Heppelman, Ladinsky Functions (FCHLF)
$\Delta^N M^{q\uparrow}_{B/\{q,q\}\uparrow:p\uparrow}(x,\vec{k}_T\cdot\vec{p}^B_T;z,p^B_{TN};Q^2)$. Correspondingly, for the final-state polarization observables we can define the Polarizing Fractured Functions (PFF)
$\Delta^N M^q_{B\uparrow/(q,q):p}(x,\vec{k}_T\cdot\vec{p}^B_T;z,p^B_{TN};Q^2)$ and the Fractured Boer Mulders Functions (FBMF)
$\Delta^N M^{q\uparrow}_{B\uparrow/\{q,q\}\uparrow:p}(x,k_{TN};z,\vec{k}_T\cdot\vec{p}_T;Q^2)$. The content of the expressions for these functions is dense and the labeling complicated so the names are attached to connect these transverse spin-odd fractured functions involving quantum diquark structures to the classification of transverse-spin odd distributions and fragmentation functions for quarks created by Mulders and Tangerman. [4] The four fractured functions presented here describe the spin-dependent differences of the conjoint probabilities for detecting a quark jet with kinematics defined by $x = x_{bj} = Q^2/2p.q$ and transverse momentum $\vec{k}_T$ and a final-state baryon with kinematics defined by the Feynman variable $z = z_B = p\cdot p_B / p\cdot q$ and transverse momentum $\vec{p}_T$ in the same deep-inelastic scattering event. A more complete discussion of these fractured functions can be found in references [1,2]. For FOF and FCHLF the $\Delta^N$ symbol indicates a target transverse spin asymmetry,

$$\Delta^N M^q_{B/(,):p\uparrow} = M^q_{B/(,):p\uparrow} - M^q_{B/(,):p\downarrow}. \tag{1.3}$$

For the fractured functions PFF and FBMF the symbol $\Delta^N$ defines a final-state polarization asymmetry from an unpolarized target,

$$\Delta^N M^q_{B\uparrow/(,):p} = M^q_{B\uparrow/(,):p} - M^q_{B\downarrow/(,):p}. \tag{1.4}$$

The set of conjoint probability densities $[M^q_{B\uparrow\downarrow/(,):p}, M^q_{B/(,):p\uparrow\downarrow},\dots]$ can capture the full range of quantum information contained in the measurements as constrained in the Bell's inequalities [5,6] and used in the study of quantum entanglement. This connection is important when viewed in the context of the powerful quantum-field-theoretical superselection principles applicable in the measurement of single-spin asymmetries involving the hard scattering of light quarks. [7,8]. For valence SIDIS these arguments also introduce quantum entanglement connecting the target fragmentation region to information in current fragmentation.

All single-spin asymmetries are odd under an operator, $O$, such that
$$O\{\vec{k}_i;\vec{\sigma}_j\}O^{-1} = \{\vec{k}_i;-\vec{\sigma}_j\}, \tag{1.5}$$
where $\vec{k}_i$ are momentum 3-vectors and $\vec{\sigma}_j$ are spin axial vectors. This operator can be seen to be the 3-D Hodge dual of the parity operator, P,
$$P\{\vec{k}_i;\vec{\sigma}_j\}P^{-1} = \{-\vec{k}_i;\vec{\sigma}_j\}. \tag{1.6}$$
The operator product $A_\tau = PO$ then has the action
$$A_\tau\{\vec{k}_i;\vec{\sigma}_j\}A_\tau^{-1} = \{-\vec{k}_i;-\vec{\sigma}_j\}. \tag{1.7}$$

The operator $A_\tau$ now designated "naïve time reversal" [9,10] is odd for transverse, parity conserving single-spin asymmetries. In this note we will denote the $A_\tau$-odd spin-directed momentum,

$$k_{TN} = \vec{k}_T \cdot (\hat{\sigma} \times \hat{Q}), \tag{1.8}$$

as the observable that defines a single-spin observable in SIDIS and other hard-scattering processes.

The explicit connection of the spin-directed momentum shift that can be produced by one unit of $\hbar$ and the asymmetries defined in Eq. (1.1) is indicated in the sketches of Fig. 1

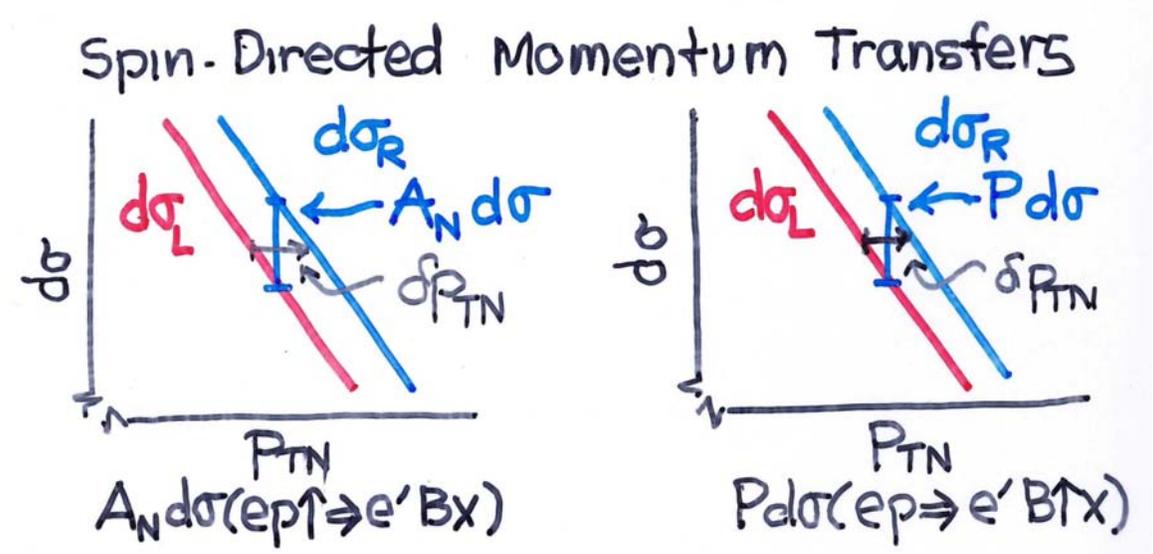

Fig. 1 Spin-directed momentum transfers provide an alternate, rigorous, definition of single-spin asymmetries.

In the discussion below we will designate a spin-directed momentum shift generated within the target nucleon by $\langle \delta k_{TN}(x,\mu^2) \rangle$ and shift generated within a QCD jet fragmentation by $\langle \delta p_{TN}(x,\mu^2) \rangle$ and treat them as directly observable asymmetries. The observation that transverse single-spin asymmetries are prohibited as $m_q \to 0$ in hard-scattering processes of perturbative QCD [11] can be designated Kane-Pumplin-Repko [12] factorization. For processes such as SIDIS this allows the use of field-theoretical idempotent projection operators,

$$\Pi_A^\pm = \left( \frac{1 \pm A_\tau}{2} \right). \tag{1.9}$$

These operators create superselection rules that isolate the spin-directed momentum shift. In processes involving the hard scattering of massless quarks $A_\tau$-odd dynamical effects can only be produced in the wave function of the target nucleon or in the color re-arrangement that incurs in the fragmentation dynamics leading to final-state hadrons. The superselection rules allow for the calculation of these shifts within the nucleon by lattice gauge simulations [13]. The application of these operators to a valence-quark SIDIS process is indicated in Fig. 2,

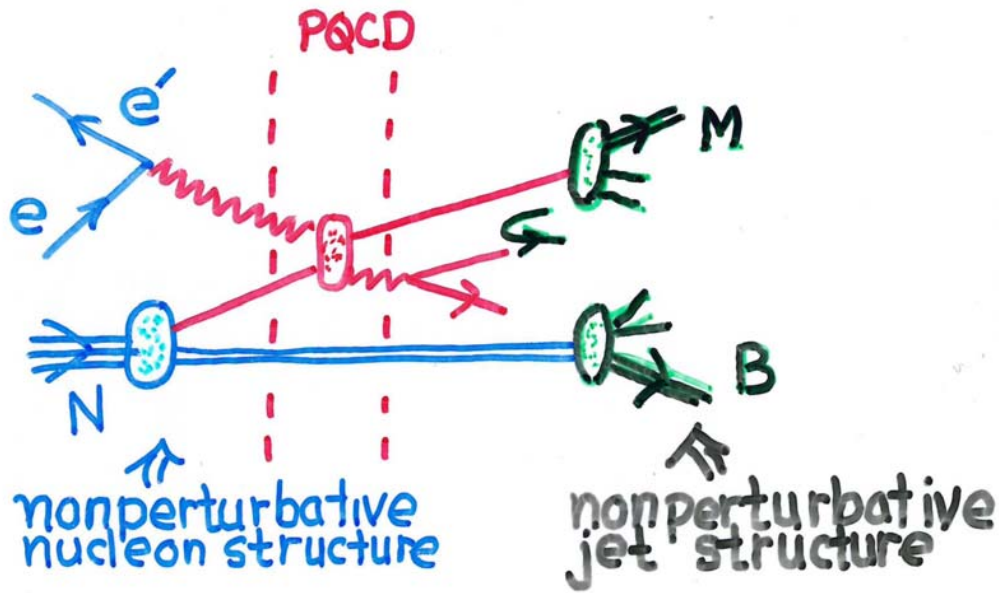

Fig. 2 The hard scattering in valence SIDIS separates nonperturbative jet structure from nonperturbative dynamics in the target nucleon.

The hard scattering process $eq \Rightarrow eqX$ in the central part of this diagram that is calculable in QCD perturbation theory for a valence u or d quark is necessarily $A_\tau$-even so that the $A_\tau$-odd dynamics leading to the spin-directed momentum shift must appear in the wave function of the target nucleon or in the color-rearrangement of the fragmentation processes leading to final-state hadrons. The phenomenological study of spin-directed momentum transfers in the current fragmentation region of this diagram [14,15] has already proved to be very instructive in isolating interesting dynamics. The diagram in Fig. 2 does not explicitly show the final-state interaction between the struck quark and the diquark structure that creates the process dependence of $A_\tau$-odd distribution functions. The quantum entanglement produced by these final-state interactions can be isolated by the projection operator $\Pi_A^-$ of (1.9) The fact that the SIDIS process demonstrates TMD factorization [16] combined with the fact that SIDIS kinematics allows for a distinction between $A_\tau$-odd dynamics occurring in the target nucleon from $A_\tau$-odd effects in fragmentation. [17] The same kinematic restrictions occur for the production of baryons in the target fragmentation region of SIDIS as can be demonstrated by the three intersecting planes shown in Fig. 3

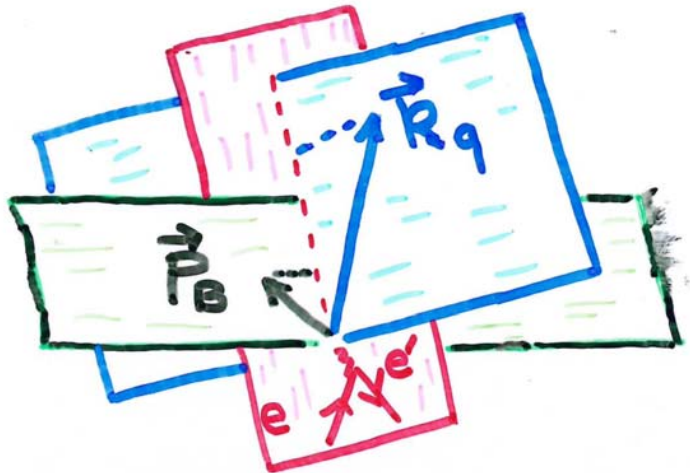

Fig. 3 The three independent planes in SIDIS can be used to differentiate between $A_\tau$-odd dynamics in the nucleon and in jet fragmentation.

A significant feature of spin-observables involving baryons is that polarization asymmetries and target spin asymmetries can be studied in tandem. For example the comparison of the FOF $\Delta^N M^u_{\Lambda/[u,d];p\uparrow}$ with the PFF $\Delta^N M^u_{\Lambda\uparrow/[u,d];p}$ in the same range of kinematics allows for the calibration of the $\langle \delta k_{TN}(x,\mu^2) \rangle$ associated with the Fractured Orbital Function with the $\langle \delta p_{TN}(z,\mu^2) \rangle$ generated in the Polarizing Fractured Function. Since these momentum shifts do not evolve under TMD evolution [11,18], the data collected from spin asymmetries can be used to study the evolution of spin-averaged TMD's.

     Most of the familiar examples of quantum entanglement involve spin observables measured in widely separated kinematical regions. [5,19] The hard scattering of a valence quark in a polarized proton as pictured in Fig. 2 can be seen to resolve the structure of a complicated internal quantum state. The Fractured Orbital Function, FOF, resolves this state in terms of an orbiting configuration involving quark-diquark and virtual meson degrees of freedom confined within the polarized proton. The final-state interactions of the struck quark that lead to a spin-directed momentum transfer involve the attractive chromo-electric force between the struck quark and the diquark spectator. This force leads to a fraction of the transverse momentum shift of the quark to be transmitted to the diquark,

$$\left\langle \delta k_{TN}(x,\mu^2)\right\rangle^{(O)}_{\{u,d\}} + \left\langle \delta k_{TN}(x,\mu^2)\right\rangle^{(O)}_{[u,d]} = -2\eta_u \left\langle \delta k_{TN}(x,\mu^2)\right\rangle^{(O)}_{u}$$
$$\left\langle \delta k_{TN}(x,\mu^2)\right\rangle^{(O)}_{\{u,u\}} = -\eta_d \left\langle \delta k_{TN}(x,\mu^2)\right\rangle^{(O)}_{d}.$$
(1.10)

In these equations the parameters $0 \leq \eta_u, \eta_d \leq 1$ provide a measure of the interaction of the quark-diquark system with the remaining internal constituents. These equations allow the prediction of $\Delta^N M^u_{p/[u,d]:p\uparrow}, \Delta^N M^u_{n/[u,d]:p\uparrow}$ and $\Delta^N M^u_{\Lambda/[u,d]:p\uparrow}$ from existing data on pion spin asymmetries. This type of long-range correlation tests the underlying understanding of orbital distributions in the same manner that the Collins conjugation prediction [20] relating the asymmetry in SIDIS and the Drell-Yan process,

$$\left\langle \delta k_{TN}(x,\mu^2)\right\rangle^{(O)}_{SIDIS} = -\left\langle \delta k_{TN}(x,\mu^2)\right\rangle^{(O)}_{DY},$$
(1.11)

when expressed in terms of spin-directed momentum transfers.

The spin state of a diquark by itself represents a quantum-entangled system and the rank dependence of the fragmentation process in the FCHLF $\Delta^N M^{q\uparrow}_{B/\{q,q'\}\uparrow:p\uparrow}$ is closely related to that of the Collins Function, $\Delta^N D_{M/q\uparrow}$. That is to say, the $A_\tau$-odd flux rupture leading to $\{q,q'\}\uparrow \Rightarrow B + \bar{q}\uparrow$ necessarily involves

$$\left\langle \delta p_{TN}(z,\mu^2)\right\rangle_{\bar{q}\uparrow} = -\left\langle \delta p_{TN}(z,\mu^2)\right\rangle_{B}.$$
(1.12)

This means that there will be spin-asymmetries associated with the production of mesons within the target fragmentation region as well as baryons. We will not discuss here the formulation of dihadron fractured functions that can be used to study spin-oriented correlations between hadrons in a full range of kinematics. At this conference Harut Avakian presented Jefferson Lab data on spin-dependent correlations between baryons and mesons. [21] I believe these data support the basic structure of the formalism presented here.

The author is grateful for informative talks and advice from Gary Goldstein and Simonetta Liutti.

## REFERENCES

1. D. Sivers, Phys. Rev. **D79**, 085008 (2009).
2. D. Sivers, Phys. Rev. **D80**, 034029 (2010).
3. See, for example, M Battaglieri et al. Acta Phys. Polon. **B46**, 2, 257 (2015).
4. R.D. Tangerman and P.J. Mulders, Phys. Rev. **D51**, 3357 (1995); Nucl. Phys. **B461**, 197 (1996).
5. J.S. Bell, Physics **1**, 195 (1964); Rev. Mod. Phys. **38**, 447 (1966).
6. P.H. Eberhard, Nuovo Cim. **B38** 75 (1977).
7. D. Sivers, Phys. Rev. **D74**, 094008 (2006).
8. D. Sivers, arXiV:0704.1791 (hep-ph) unpublished
9. R. Jaffe, in *Proceedings on the Workshop on Deep Inelastic Scattering off Polarized Targets, DEZY-Zeuthen, Germany 1997,* edited by J.Blumlein et al., 167.

10. D. Sivers, in *Proceedings of the Workshop on Deep Inelastic Scattering off Polarized Targets, DEZY-Zeuthen, Germany 1997,* edited by J. Blumlein et al., 383.
11. D. Sivers, Int. J. Mod. Phys. Conf. Ser. **25**, 1460002 (2014).
12. G. Kane, J. Pumplin and W. Repko, Phys. Rev. Lett. **41**, 1689 (1978).
13. B. U. Musch, et al., Phys. Rev. **D85**, 094510 (2012).
14. M. Anselmino et al. , arXiV:1107.4446 (2011).
15. Z.-B. Kang, et al., arXiV:1505.05589 (2015).
16. J. Collins, arXiV:1307.2920 (2013).
17. A. Bacchetta et al., Phys. Rev. **D70**, 117504 (2011).
18. D. Sivers, in EBJ Web. Conf. **85**, 2007 (2015).
19. A. Einstein, B Podolsky and N. Rosen, Phys. Rev. **D47**, 777 (1935).
20. J. Collins, Phys. Lett. **B536**, 43 (2002).
21. H. Avakian, this conference.